\documentstyle[aps,preprint]{revtex}

\begin{document}
\author{Remo Garattini}
\address{M\'{e}canique et Gravitation, Universit\'{e} de Mons-Hainaut,\\
Facult\'e des Sciences, 15 Avenue Maistriau, \\
B-7000 Mons, Belgium \\
and\\
Facolt\`a di Ingegneria, Universit\`a degli Studi di Bergamo,\\
Viale Marconi, 5, 24044 Dalmine (Bergamo) Italy\\
e-mail: Garattini@mi.infn.it}
\title{Vacuum Energy, Variational Methods and the Casimir Energy}
\date{\today}
\maketitle

\begin{abstract}
Following the subtraction procedure for manifolds with boundaries, we
calculate by variational methods, the Schwarzschild and Flat space energy
difference. The one loop approximation for TT tensors is considered here. An
analogy between the computed energy difference in momentum space and the
Casimir effect is illustrated. We find a singular behaviour in the UV-limit,
due to the presence of the horizon when $r=2m.$ When $r>2m$ this singular
behaviour disappears, which is in agreement with various other models
previously presented.
\end{abstract}

\section{Introduction}

An interesting problem appearing in Einstein gravity is the computation of
quantum corrections to a classical energy. A possible approach is the
analysis of the thermodynamical quantities that characterize the system
under consideration. This analysis can be carried out by computing the
system free energy at a given volume and temperature, by means of a
partition function and the Euclidean action. Following the background
method, we fix a metric and look at quantum fluctuations with respect to
such a background with the appropriate boundary conditions, then we
functionally integrate such metric fluctuations which are strictly periodic
in Euclidean time $t$. In particular, the only feasible way to treat
functional integration is by saddle-point methods. This is adequate for the
treatment of the small perturbation concerning Minkowski space and for a
semiclassical analysis of vacuum stability. However, a different point of
view based on the Hamiltonian approach, could be considered. In this
framework, quantum corrections to a classical energy can be computed by
means of expectation values of the total Hamiltonian with respect to some
states. It is clear that the problem is too large to be completely solved.
To this end we might take into consideration the simplest non trivial
saddle-point we can extract from vacuum Einstein equations, the
Schwarzschild solution 
\begin{equation}
ds^{2}=-\left( 1-\frac{2MG}{r}\right) dt^{2}+\left( 1-\frac{2MG}{r}\right)
^{-1}dr^{2}+r^{2}d\Omega ^{2},
\end{equation}
where $d\Omega ^{2}=d\theta ^{2}+\sin ^{2}\theta d\phi ^{2}$ is the line
element of the unit sphere, $G$ is Newton's constant and $M$ is a parameter
representing the mass of the wormhole. This metric is asymptotically flat.
The apparent singularity located at $r=2MG$ can be removed by a suitable
definition of the coordinates, e.g., the Kruskal-Szekeres coordinates, which
is written as 
\begin{equation}
\left\{ 
\begin{array}{c}
\left( \frac{r}{2MG}-1\right) \exp \left( \frac{r}{2MG}\right) =xy \\ 
\exp \left( \frac{t}{2MG}\right) =\frac{x}{y}
\end{array}
\right. .
\end{equation}
In terms of these coordinates we have 
\begin{equation}
ds^{2}=\frac{32\left( MG\right) ^{3}}{r}\exp \left( -\frac{r}{2MG}\right)
dxdy+r^{2}d\Omega ^{2}.
\end{equation}
The only true singularities are at curves $xy=-1$, where $r=0$. The region $%
\left\{ x>0,y>0\right\} $ is the ``{\it outside region}'', the only region
from which distant observers can obtain any information. The line $y=0$,
where $r=2MG$, is the ``{\it future horizon}''; the line $x=0$ where also $%
r=2MG$, is the ``{\it past horizon}''. We will consider a slice $\Sigma $ of
the Schwarzschild manifold representing a constant time section of ${\cal M}$%
. This surface $\Sigma $ is an Einstein-Rosen bridge with wormhole topology $%
S^{2}\times R^{1}$ which defines a bifurcation surface, dividing $\Sigma $
in two parts denoted by $\Sigma _{+}$ and $\Sigma _{-}$. Our purpose is to
consider perturbations at $\Sigma $ with $t$ constant, which naturally
define quantum fluctuations of the Einstein-Rosen bridge. In particular we
will focus our attention on the $\Sigma _{+}$ sector of the manifold,
corresponding to the ``{\it outside region}'' of the Kruskal manifold. The
explicit expression of the Hamiltonian can be calculated by considering the
line element 
\begin{equation}
ds_{{}}^{2}=-N^{2}\left( dx^{0}\right) ^{2}+g_{ij}\left(
N^{i}dx^{0}+dx^{i}\right) \left( N^{j}dx^{0}+dx^{j}\right) ,  \label{i00}
\end{equation}
where $N$ is called the {\it lapse} function and $N_{i}$ is the {\it shift }%
function. When $N=\sqrt{1-\frac{2MG}{r}}$, $N_{i}=0$ and $%
g_{ij}dx^{i}dx^{j}=\left( 1-\frac{2MG}{r}\right) ^{-1}dr^{2}+r^{2}d\Omega
^{2}$, we recover the Schwarzschild solution. On the slice $\Sigma $,
deviations from the Schwarzschild metric spatial section will be considered 
\begin{equation}
g_{ij}=\bar{g}_{ij}+h_{ij},  \label{i0}
\end{equation}
with $N_{i}=0$ and $N\equiv N\left( r\right) $. Then the line element $%
\left( \ref{i00}\right) $ becomes 
\begin{equation}
ds_{{}}^{2}=-N^{2}\left( r\right) \left( dx^{0}\right)
^{2}+g_{ij}dx^{i}dx^{j}
\end{equation}
and the total Hamiltonian is 
\begin{equation}
H_{T}=H_{\Sigma }+H_{\partial \Sigma }=\int_{\Sigma }d^{3}x(N{\cal H+}N_{i}%
{\cal H}^{i})+H_{\partial \Sigma },
\end{equation}
where 
\begin{equation}
{\cal H}{\bf =}G_{ijkl}\pi ^{ij}\pi ^{kl}\left( \frac{l_{p}^{2}}{\sqrt{g}}%
\right) -\left( \frac{\sqrt{g}}{l_{p}^{2}}\right) R^{\left( 3\right) }\ 
\text{ (Super Hamiltonian),}
\end{equation}
\begin{equation}
{\cal H}^{i}=-2\pi _{|j}^{ij}\ \text{(Super Momentum),}
\end{equation}
while $H_{\partial \Sigma }$ represents the energy stored in the boundaries.
In this respect, we will follow the ${\cal ADM}$ approach\cite{ADM}, even
though the quasilocal energy context gives a more general treatment with the
possibility of looking at the gravitational thermodynamics\cite
{BroYor,FroMar}. Moreover, since the space under investigation is
asymptotically flat in spacelike directions, the quasilocal energy agrees
with the results of the ${\cal ADM}$ approach in the limit that the boundary
tends to spatial infinity. In any case, to correctly compute $H_{\partial
\Sigma }=H_{{\cal ADM}}^{{}}$ we have to fix a reference frame to normalize
the energy value on the boundary to zero. This opens the problem of the
subtraction procedure investigated in Refs.\cite{FroMar,HawHor}. In this
paper we would like to apply such a procedure extended to the volume term,
at least at one loop. Since the reference space for the Schwarzschild metric
is flat space, the contribution to the energy term is 
\begin{equation}
H_{{\cal ADM}}=\lim_{r\rightarrow \infty }\int_{\partial \Sigma }\sqrt{%
\widehat{g}}\widehat{g}^{ij}\left[ \widehat{g}_{ik,j}-\widehat{g}_{ij,k}%
\right] dS^{k}=M
\end{equation}
where $\widehat{g}_{ij}$ is the metric induced on a spacelike hypersurface $%
\partial \Sigma $ which has a boundary at infinity like $S^{2}$. Following
the result of Ref.\cite{HawHor}, we see that $H_{{\cal ADM}}$ is completely
equivalent to 
\begin{equation}
-\frac{1}{8\pi G}\int_{\partial \Sigma }\left[ ^{2}K-^{2}K_{0}\right] ,
\end{equation}
where the subtraction structure is evident. The one-loop contribution to the
zero point energy for gravitons embedded in flat space is 
\begin{equation}
2\cdot \frac{1}{2}\int \frac{d^{3}k}{\left( 2\pi \right) ^{3}}\sqrt{k^{2}}.
\end{equation}
It is clear that this term is UV\ divergent. We will show that the same kind
of divergence is present when the curved background is considered. In the
spirit of the subtraction procedure we will compute the difference between
zero point energies. Their difference at one loop represents a Casimir-like
computation. The paper is structured as follows, in section \ref{p2} we
define the gaussian wave functional for gravity and we analyze the
orthogonal decomposition of the metric deformations, in section \ref{p3} we
give some of the basic rules to perform the functional integration and we
define the Hamiltonian approximated up to second order, in section \ref{p4},
we analyze the spin-two operator acting on transverse traceless tensors,
only for positive values of $E^{2}$. We summarize and conclude in section 
\ref{p5}.

\section{ Energy Density Calculation in Schr\"{o}dinger Representation}

\label{p2}As already mentioned, we would like to discuss the possibility of
generalizing the boundary subtraction procedure. To this end, by looking at
the Hamiltonian structure, we see that there are two classical constraints 
\begin{equation}
\left\{ 
\begin{array}{l}
{\cal H}\text{ }=0 \\ 
{\cal H}^{i}=0
\end{array}
\right. ,
\end{equation}
which are satisfied both by the Schwarzschild and Flat metric and two {\it %
quantum} constraints 
\begin{equation}
\left\{ 
\begin{array}{l}
{\cal H}\tilde{\Psi}\text{ }=0 \\ 
{\cal H}^{i}\tilde{\Psi}=0
\end{array}
\right. .
\end{equation}
${\cal H}\tilde{\Psi}$ $=0$ is known as the {\it Wheeler-DeWitt} equation
(WDW). Nevertheless, we are interested in assigning a meaning to 
\begin{equation}
\frac{\left\langle \Psi \left| H_{\Sigma }^{Schw.}-H_{\Sigma }^{Flat}\right|
\Psi \right\rangle }{\left\langle \Psi |\Psi \right\rangle }+\frac{%
\left\langle \Psi \left| H_{{\cal ADM}}\right| \Psi \right\rangle }{%
\left\langle \Psi |\Psi \right\rangle },  \label{a1}
\end{equation}
where $\Psi $ is a wave functional whose structure will be determined later
and $H_{\Sigma }^{Schw.}\left( H_{\Sigma }^{Flat}\right) $ is the total
Hamiltonian referred to the different spacetimes for the volume term. This
has to be meant in this way: it is true that the WDW equation refers to the
space of metrics, but the space of metrics posses different sectors\cite
{ReggeTeitelboim} and we are considering the sector of asymptotically flat
metrics, in which the zero point energy is defined with respect to Minkowski
space. For the de Sitter sector, we have to substract the energy of de
Sitter background and so on. Note that if the expectation value is
calculated on the wave functional solution of the WDW equation, we obtain
only the boundary contribution. However, in this context boundaries are at
infinity in spacelike directions, that it is equivalent to considering the
unphysical situation of computing energy excitation in the asymptotic
region. Then to give meaning to $\left( \ref{a1}\right) $, we adopt the
semiclassical strategy of the WKB expansion. By observing that the kinetic
part of the Super Hamiltonian is quadratic in the momenta, we expand the
three-scalar curvature $\int d^{3}x\sqrt{g}R^{\left( 3\right) }$ up to $%
o\left( h^{3}\right) $ and we get 
\begin{equation}
\int d_{{}}^{3}x\left[ -\frac{1}{4}h\triangle h+\frac{1}{4}h^{li}\triangle
h_{li}-\frac{1}{2}h^{ij}\nabla _{l}\nabla _{i}h_{j}^{l}+\frac{1}{2}h\nabla
_{l}\nabla _{i}h_{{}}^{li}-\frac{1}{2}h^{ij}R_{ia}h_{j}^{a}+\frac{1}{2}%
hR_{ij}h_{{}}^{ij}\right] ,
\end{equation}
where $h$ is the trace of $h_{ij}$. On the other hand, following the usual
WKB expansion, we will consider $\tilde{\Psi}\simeq C\exp \left( iS\right) $%
. In this context, the approximated wave functional will be substituted by a 
{\it trial wave functional} according to the variational approach we would
like to implement as regards this problem.

\section{ The Gaussian Wave Functional for TT tensors}

\label{p3}

To actually make such calculations, we need an orthogonal decomposition for
both $\pi _{ij\text{ }}^{}$and $h_{ij}^{}$ to disentangle gauge modes from
physical deformations. We define the inner product

\begin{equation}
\left\langle h,k\right\rangle :=\int_{{\cal M}}^{}\sqrt{g}%
G^{ijkl}h_{ij}\left( x\right) k_{kl}\left( x\right) d^3x,
\end{equation}
by means of the inverse WDW metric $G_{ijkl}$, to have a metric on the space
of deformations, i.e. a quadratic form on the tangent space at h, with

\begin{equation}
\begin{array}{c}
G^{ijkl}=(g^{ik}g^{jl}+g^{il}g^{jk}-2g^{ij}g^{kl})\text{.}
\end{array}
\end{equation}
The inverse metric is defined on co-tangent space and it assumes the form

\begin{equation}
\left\langle p,q\right\rangle :=\int_{{\cal M}}^{}\sqrt{g}%
G_{ijkl}p^{ij}\left( x\right) q^{kl}\left( x\right) d^3x\text{,}
\end{equation}
so that

\begin{equation}
G^{ijnm}G_{nmkl}=\frac 12\left( \delta _k^i\delta _l^j+\delta _l^i\delta
_k^j\right) .
\end{equation}
Note that in this scheme the ``inverse metric'' is actually the WDW metric
defined on phase space. Now, we have the desired decomposition on the
tangent space of 3-metric deformations\cite{BergerEbin,York}:

\begin{equation}
h_{ij}=\frac 13hg_{ij}+\left( L\xi \right) _{ij}+h_{ij}^{\bot }  \label{b0}
\end{equation}
where the operator $L$ maps $\xi _i$ into symmetric tracefree tensors

\begin{equation}
\left( L\xi \right) _{ij}=\nabla _i\xi _j+\nabla _j\xi _i-\frac 23%
g_{ij}\left( \nabla \cdot \xi \right) .
\end{equation}
Then the inner product between three-geometries becomes 
\[
\left\langle h,h\right\rangle :=\int_{{\cal M}}\sqrt{g}G^{ijkl}h_{ij}\left(
x\right) h_{kl}\left( x\right) d^3x= 
\]
\begin{equation}
\int_{{\cal M}}\sqrt{g}\left[ -\frac 23h^2+\left( L\xi \right) ^{ij}\left(
L\xi \right) _{ij}+h^{ij\bot }h_{ij}^{\bot }\right] .  \label{b1}
\end{equation}
With the orthogonal decomposition in hand we can define a ``{\it Vacuum
Trial State}'' 
\begin{equation}
\Psi \left[ h_{ij}\left( \overrightarrow{x}\right) \right] ={\cal N}\exp
\left\{ -\frac 1{4l_p^2}\left[ \left\langle hK^{-1}h\right\rangle
_{x,y}^{\bot }+\left\langle \left( L\xi \right) K^{-1}\left( L\xi \right)
\right\rangle _{x,y}^{\Vert }+\left\langle hK^{-1}h\right\rangle
_{x,y}^{Trace}\right] \right\} ,
\end{equation}
which will be used as a probe for the gravitational ground state. This
particular expression is useful because the functional can be represented as
a product of three functionals defined on the decomposed tensor field 
\begin{equation}
\Psi \left[ h_{ij}\left( \overrightarrow{x}\right) \right] ={\cal N}\Psi %
\left[ h_{ij}^{\bot }\left( \overrightarrow{x}\right) \right] \Psi \left[
\left( L\xi \right) _{ij}\right] \Psi \left[ \frac 13g_{ij}h\left( 
\overrightarrow{x}\right) \right] .  \label{b1a}
\end{equation}
$h_{ij}^{\bot }$ is the tracefree-transverse part of the $3D$ quantum field, 
$\left( L\xi \right) _{ij}$ is the longitudinal part and finally $h$ is the
trace part of the same field. $\left\langle \cdot ,\cdot \right\rangle
_{x,y} $ denotes space integration and $K^{-1}$ is the inverse propagator
containing variational parameters. The main reason for a similar ``{\it %
Ansatz}'' comes from the observation that the quadratic part in the momenta
of the Hamiltonian decouples in the same way of eq.$\left( \ref{b1}\right) $%
. Note that the decomposition related to the momenta is independent of the
choice of the functional. To calculate the energy density, we need to know
the action of some basic operators on $\Psi \left[ h_{ij}\right] $. The
action of the operator $h_{ij}$ on $|\Psi \rangle =\Psi \left[ h_{ij}\right] 
$ is realized by 
\begin{equation}
h_{ij}\left( x\right) |\Psi \rangle =h_{ij}\left( \overrightarrow{x}\right)
\Psi \left[ h_{ij}\right] .
\end{equation}
The action of the operator $\pi _{ij}$ on $|\Psi \rangle $, in general, is

\begin{equation}
\pi _{ij}\left( x\right) |\Psi \rangle =-i\frac{\delta }{\delta h_{ij}\left( 
\overrightarrow{x}\right) }\Psi \left[ h_{ij}\right] .
\end{equation}
The inner product is defined by the functional integration: 
\begin{equation}
\left\langle \Psi _{1}\mid \Psi _{2}\right\rangle =\int \left[ {\cal D}h_{ij}%
\right] \Psi _{1}^{\ast }\left\{ h_{ij}\right\} \Psi _{2}\left\{
h_{kl}\right\} ,
\end{equation}
and the energy eigenstates satisfy the stationary Schr\"{o}dinger equation: 
\begin{equation}
\int d^{3}x{\cal H}\left\{ -i\frac{\delta }{\delta h_{ij}\left( 
\overrightarrow{x}\right) },h_{ij}\left( \overrightarrow{x}\right) \right\}
\Psi \left\{ h_{ij}\right\} =E\Psi \left\{ h_{ij}\right\} ,  \label{b2}
\end{equation}
where ${\cal H}\left\{ -i\frac{\delta }{\delta h_{ij}\left( x\right) }%
,h_{ij}\left( x\right) \right\} $ is the Hamiltonian density. Note that the
previous equation in the general context of Einstein gravity is devoid of
meaning, because of the constraints. However in the semiclassical context,
we can give a meaning to eq.$\left( \ref{b2}\right) $, where a {\it %
semiclassical time} is introduced in the same manner of Refs.\cite
{HalHaw,Halliwell}. There, a Schr\"{o}dinger equation of the form 
\begin{equation}
i\frac{\partial \Psi ^{\bot }}{\partial t}=H_{|2}\Psi ^{\bot }  \label{b3a}
\end{equation}
is recovered by the WDW equation approximated to second order for a
perturbed minisuperspace Friedmann model without boundary terms. When
asymptotically flat boundary terms are present we have to take account of
such contributions in the WKB expansion such as in Ref.\cite{Brotz}. However
in this paper only gravitational transverse-traceless modes are considered
on the fixed curved background and $\Psi ^{\bot }$ is substituted by a trial
wave functional. To further proceed, instead of solving $\left( \ref{b2}%
\right) $, which is of course impossible, we can formulate the same problem
by means of a variational principle. We demand that 
\begin{equation}
\frac{\left\langle \Psi \mid H\mid \Psi \right\rangle }{\left\langle \Psi
\mid \Psi \right\rangle }=\frac{\int \left[ {\cal D}g_{ij}^{\bot }\right]
\int d_{{}}^{3}x\Psi _{1}^{\ast }\left\{ g_{ij}^{\bot }\right\} {\cal H}\Psi
\left\{ g_{kl}^{\bot }\right\} }{\int \left[ {\cal D}g_{ij}^{\bot }\right]
\mid \Psi \left\{ g_{ij}^{\bot }\right\} \mid ^{2}}  \label{b2a}
\end{equation}
be stationary against arbitrary variations of $\Psi \left\{ h_{ij}\right\} $%
. The form of $\left\langle \Psi \mid H\mid \Psi \right\rangle $ can be
computed as follows. We define normalized mean values 
\begin{equation}
\bar{g}_{ij}^{\bot }\left( \overrightarrow{x}\right) =\frac{\int \left[ 
{\cal D}g_{ij}^{\bot }\right] \int d_{{}}^{3}xg_{ij}^{\bot }\left( 
\overrightarrow{x}\right) \mid \Psi \left\{ g_{ij}^{\bot }\right\} \mid ^{2}%
}{\int \left[ {\cal D}g_{ij}^{\bot }\right] \mid \Psi \left\{ g_{ij}^{\bot
}\right\} \mid ^{2}},
\end{equation}
\begin{equation}
\bar{g}_{ij}^{\bot }\left( \overrightarrow{x}\right) \text{ }\bar{g}%
_{kl}^{\bot }\left( \overrightarrow{y}\right) +K_{ijkl^{{}}}^{\bot }\left( 
\overrightarrow{x},\overrightarrow{y}\right) 
\end{equation}
\begin{equation}
=\frac{\int \left[ {\cal D}g_{ij}^{\bot }\right] \int
d_{{}}^{3}xg_{ij}^{\bot }\left( \overrightarrow{x}\right) g_{kl}^{\bot
}\left( \overrightarrow{y}\right) \mid \Psi \left\{ g_{ij}^{\bot }\right\}
\mid ^{2}}{\int \left[ {\cal D}g_{ij}^{\bot }\right] \mid \Psi \left\{
g_{ij}^{\bot }\right\} \mid ^{2}}.
\end{equation}
It follows that, by defining $h_{ij}^{\bot }=g_{ij}-\bar{g}_{ij}$, we have 
\begin{equation}
\int \left[ {\cal D}h_{ij}^{\bot }\right] h_{ij}^{\bot }\left( 
\overrightarrow{x}\right) \mid \Psi \left\{ h_{ij}^{\bot }+\bar{g}%
_{ij}^{\bot }\right\} \mid ^{2}=0  \label{b3}
\end{equation}
and 
\[
\int \left[ {\cal D}h_{ij}^{\bot }\right] \int d_{{}}^{3}xh_{ij}^{\bot
}\left( \overrightarrow{x}\right) h_{kl}^{\bot }\left( \overrightarrow{y}%
\right) \mid \Psi \left\{ h_{ij}^{\bot }+\bar{g}_{ij}^{\bot }\right\} \mid
^{2}=
\]
\begin{equation}
K_{ijkl^{{}}}^{\bot }\left( \overrightarrow{x},\overrightarrow{y}\right)
\int \left[ {\cal D}h_{ij}^{\bot }\right] \mid \Psi \left\{ h_{ij}^{\bot }+%
\bar{g}_{ij}^{\bot }\right\} \mid ^{2}.  \label{b4}
\end{equation}
Nevertheless, the application of the variational principal on arbitrary wave
functional does not improve the situation described by the eq.$\left( \ref
{b2}\right) $. To this purpose, we give to the trial wave functional the
form 
\begin{equation}
\Psi \left[ h_{ij}^{\bot }\right] ={\cal N}\exp \left\{ -\frac{1}{4l_{p}^{2}}%
\left\langle \left( g-\overline{g}\right) K^{-1}\left( g-\overline{g}\right)
\right\rangle _{x,y}^{\bot }\right\} .  \label{b4a}
\end{equation}
We immediately conclude that 
\begin{equation}
\left\langle \Psi |\pi _{ij}^{\bot }\left( \overrightarrow{x}\right) |\Psi
\right\rangle =0
\end{equation}
where $\pi _{ij}^{\bot }$ is the TT momentum. In Appendix \ref{p7}, we will
show that 
\begin{equation}
\left\langle \Psi |\pi _{ij}^{\bot }\left( \overrightarrow{x}\right) \pi
_{kl}^{\bot }\left( \overrightarrow{y}\right) |\Psi \right\rangle =\frac{1}{4%
}K_{ijkl}^{-1}\left( \overrightarrow{x},\overrightarrow{y}\right) .
\label{b5}
\end{equation}
Choice $\left( \ref{b4a}\right) $ is related to the form of the Hamiltonian
approximated to quadratic order in the metric deformations. Indeed, up to
this order we have a harmonic oscillator whose ground state has a Gaussian
form. By means of decomposition $\left( \ref{b0}\right) $, we extract the TT
sector contribution in the previous expression. Moreover, the functional
representation $\left( \ref{b1a}\right) $ eliminates every interaction
between gauge and the other terms. Then for the TT sector (spin-two), one
gets 
\begin{equation}
\int_{\Sigma }d^{3}x\sqrt{g}R^{\left( 3\right) }\simeq \frac{1}{4l_{p}^{2}}%
\int_{\Sigma }d^{3}x\sqrt{g}\left[ h^{\bot ij}\left( \triangle _{2}\right)
_{j}^{a}h_{ia}^{\bot }-2hR_{ij}h^{\bot ij}\right] ,  \label{b5a}
\end{equation}
where $\left( \triangle _{2}\right) _{j}^{a}:=-\triangle \delta
_{j}^{a}+2R_{j}^{a}$. The latter term disappears because the gaussian
integration does not mix the components. Then by collecting together eq.$%
\left( \ref{b5a}\right) $ and eq. $\left( \ref{b5}\right) $, one obtains the
one-loop-like Hamiltonian form for TT deformations 
\begin{equation}
H^{\bot }=\frac{1}{4l_{p}^{2}}\int_{{\cal M}}^{{}}d^{3}x\sqrt{g}G^{ijkl}%
\left[ K^{-1\bot }\left( x,x\right) _{ijkl}+\left( \triangle _{2}\right)
_{j}^{a}K^{\bot }\left( x,x\right) _{iakl}\right] .  \label{b6}
\end{equation}
The propagator $K^{\bot }\left( x,x\right) _{iakl}$ comes from a functional
integration and it can be represented as 
\begin{equation}
K^{\bot }\left( \overrightarrow{x},\overrightarrow{y}\right)
_{iakl}:=\sum_{N}\frac{h_{ia}^{\bot }\left( \overrightarrow{x}\right)
h_{kl}^{\bot }\left( \overrightarrow{y}\right) }{2\lambda _{N}\left(
p\right) },
\end{equation}
where $h_{ia}^{\bot }\left( \overrightarrow{x}\right) $ are the
eigenfunctions of $\triangle _{2j}^{a}$ and $\lambda _{N}\left( p\right) $
are infinite variational parameters.

\section{The Spectrum of the Spin-2 Operator and the evaluation of the
Energy Density in Momentum Space}

\label{p4}

The Spin-two operator is defined by

\begin{equation}
\left( \triangle _2\right) _j^a:=-\triangle \delta _j^{a_{}^{}}+2R_j^a
\end{equation}
where $\triangle $ is the curved Laplacian (Laplace-Beltrami operator) on a
Schwarzschild background and $R_{j\text{ }}^a$ is the mixed Ricci tensor
whose components are:

\begin{equation}
R_j^a=diag\left\{ \frac{-2m}{r_{}^3},\frac m{r_{}^3},\frac m{r_{}^3}\right\}
,
\end{equation}
where $2m=2MG$. This operator is similar to the Lichnerowicz operator
provided that we substitute the Riemann tensor with the Ricci tensor. This
is essentially due to the fact that the Riemann tensor in three-dimensions
is a linear combination of the Ricci tensor. In $\left( \ref{d1}\right) $
the Ricci tensor acts as a potential on the space of TT tensors; for this
reason we are led to study the following eigenvalue equation

\begin{equation}
\left( -\triangle \delta _{j}^{a}+2R_{j}^{a}\right)
h_{a}^{i}=E^{2}h_{j}^{i_{{}}^{{}}}  \label{d1}
\end{equation}
where $E^{2}$ is the eigenvalue of the corresponding equation. In doing so,
we follow Regge and Wheeler in analyzing the equation as modes of definite
frequency, angular momentum and parity. We can specialize to the case with
the quantum number corresponding to the projection of the angular momentum
on the z-axis is zero, without altering the contribution to the total energy
because of the spherical symmetry of the problem. In this case,
Regge-Wheeler decomposition \cite{Regge} shows that the even-parity
three-dimensional perturbation is

\begin{equation}
h_{ij}^{even}\left( r,\vartheta ,\phi \right) =diag\left[ H\left( r\right)
\left( 1-\frac{2m}{r}\right) ^{-1},r^{2}K\left( r\right) ,r^{2}\sin
^{2}\vartheta K\left( r\right) \right] Y_{l0}\left( \vartheta ,\phi \right) .
\label{d2}
\end{equation}
Representation $\left( \ref{d2}\right) $ shows a gravitational perturbation
decoupling. For a generic value of the angular momentum.angular momentum $l$%
, one gets

\begin{equation}
\left\{ 
\begin{array}{c}
-\triangle _lH\left( r\right) -\frac{4m}{r_{}^3}H\left( r\right)
=E_l^2H\left( r\right) \\ 
\\ 
-\triangle _lK\left( r\right) +\frac{2m}{r_{}^3}K\left( r\right)
=E_l^2K\left( r\right) \\ 
\\ 
-\triangle _lK\left( r\right) +\frac{2m}{r_{}^3}K\left( r\right)
=E_l^2K\left( r\right) .
\end{array}
\right.  \label{d3}
\end{equation}
The Laplacian in this particular geometry can be written as

\begin{equation}
\triangle _l=\left( 1-\frac{2m}r\right) \frac{d^2}{dr^2}+\left( \frac{2r-3m}{%
r_{}^2}\right) \frac d{dr}-\frac{l\left( l+1\right) }{r_{}^2}.
\end{equation}
Defining reduced fields, such as:

\begin{equation}
H\left( r\right) =\frac{h\left( r\right) }r;\qquad K\left( r\right) =\frac{%
k\left( r\right) }r,
\end{equation}
and changing variables to

\begin{equation}
x=2m\left\{ \sqrt{\frac r{2m}}\sqrt{\frac r{2m}-1}+\ln \left( \sqrt{\frac r{%
2m}}+\sqrt{\frac r{2m}-1}\right) \right\} ,  \label{d2a}
\end{equation}
the system $\left( \ref{d3}\right) $ becomes

\begin{equation}
\left\{ 
\begin{array}{c}
-\frac{d^2}{dx^2}h\left( x\right) +V^{-}\left( x\right) h\left( x\right)
=E_l^2h\left( x\right) \\ 
\\ 
-\frac{d^2}{dx^2}k\left( x\right) +V^{+}\left( x\right) k\left( x\right)
=E_l^2k\left( x\right) \\ 
\\ 
-\frac{d^2}{dx^2}k\left( x\right) +V^{+}\left( x\right) k\left( x\right)
=E_l^2k\left( x\right)
\end{array}
\right.  \label{d4}
\end{equation}
where 
\begin{equation}
V^{\mp }\left( x\right) =\frac{l\left( l+1\right) }{r^2\left( x\right) }\mp 
\frac{3m}{r\left( x\right) ^3}.
\end{equation}
This new variable represents the proper geodesic distance from the wormhole
throat such that 
\[
\text{when }r\longrightarrow \infty \text{, }x\simeq r\text{ \quad and }%
V^{\mp }\left( x\right) \longrightarrow 0 
\]
\begin{equation}
\text{when }r\longrightarrow r_0\text{, }x\simeq 0\text{\quad and }V^{\mp
}\left( x\right) \longrightarrow \frac{l\left( l+1\right) }{r_0^2}\mp \frac{%
3m}{r_0^3}=const,
\end{equation}
where $r_0$ satisfies the condition $r_0>2m$. The solution of $\left( \ref
{d4}\right) $, in both cases (flat and curved one) is the spherical Bessel
function of the first kind 
\begin{equation}
j_0\left( px\right) =\sqrt{\frac 2\pi }\sin \left( px\right)
\end{equation}
This choice is dictated by the requirement that 
\begin{equation}
h\left( x\right) ,k\left( x\right) \rightarrow 0\text{\qquad when\qquad }%
x\rightarrow 0\ \left( \text{alternatively }r\rightarrow 2m\right) .
\end{equation}
Then 
\begin{equation}
K\left( x,y\right) =\frac{j_0\left( px\right) j_0\left( py\right) }{2\lambda 
}\cdot \frac 1{4\pi }  \label{d5}
\end{equation}
Substituting $\left( \ref{d5}\right) $ in $\left( \ref{b6}\right) $ one gets
(after normalization in spin space and after a rescaling of the fields in
such a way as to absorb $l_p^2$) 
\begin{equation}
E\left( m,\lambda \right) =\frac V{2\pi ^2}\sum_{l=0}^\infty
\sum_{i=1}^2\int_0^\infty dpp^2\left[ \lambda _i\left( p\right) +\frac{%
E_i^2\left( p,m,l\right) }{\lambda _i\left( p\right) }\right]  \label{d6}
\end{equation}
where 
\begin{equation}
E_{1,2}^2\left( p,m,l\right) =p^2+\frac{l\left( l+1\right) }{r_0^2}\mp \frac{%
3m}{r_0^3},
\end{equation}
$\lambda _i\left( p\right) $ are variational parameters corresponding to the
eigenvalues for a (graviton) spin-two particle in an external field and $V$
is the volume of the system.

By minimizing $\left( \ref{d6}\right) $ with respect to $\lambda _i\left(
p\right) $ one obtains $\overline{\lambda }_i\left( p\right) =\left[
E_i^2\left( p,m,l\right) \right] ^{\frac 12}$ and 
\begin{equation}
E\left( m,\overline{\lambda }\right) =\frac V{2\pi ^2}\sum_{l=0}^\infty
\sum_{i=1}^2\int_0^\infty dp2\sqrt{E_i^2\left( p,m,l\right) }
\end{equation}
with 
\[
p^2+\frac{l\left( l+1\right) }{r_0^2}>\frac{3m}{r_0^3}. 
\]
Thus, in presence of the curved background, we get 
\begin{equation}
E\left( m\right) =\frac V{2\pi ^2}\frac 12\sum_{l=0}^\infty \int_0^\infty
dpp^2\left( \sqrt{p^2+c_{-}^2}+\sqrt{p^2+c_{+}^2}\right)  \label{d7}
\end{equation}
where 
\[
c_{\mp }^2=\frac{l\left( l+1\right) }{r_0^2}\mp \frac{3m}{r_0^3}, 
\]
while when we refer to the flat space, in the spirit of the subtraction
procedure, we have $m=0$ and $c^2=$ $\frac{l\left( l+1\right) }{r_0^2}$.
Then 
\begin{equation}
E\left( 0\right) =\frac V{2\pi ^2}\frac 12\sum_{l=0}^\infty \int_0^\infty
dpp^2\left( 2\sqrt{p^2+c^2}\right)  \label{d8}
\end{equation}
Now, we are in position to compute the difference between $\left( \ref{d7}%
\right) $ and $\left( \ref{d8}\right) $. Since we are interested in the $UV$
limit, we have 
\[
\Delta E\left( m\right) =E\left( m\right) -E\left( 0\right) 
\]
\[
=\frac V{2\pi ^2}\frac 12\sum_{l=0}^\infty \int_0^\infty dpp^2\left[ \sqrt{%
p^2+c_{-}^2}+\sqrt{p^2+c_{+}^2}-2\sqrt{p^2+c^2}\right] 
\]
\begin{equation}
=\frac V{2\pi ^2}\frac 12\sum_{l=0}^\infty \int_0^\infty dpp^3\left[ \sqrt{%
1+\left( \frac{c_{-}}p\right) ^2}+\sqrt{1+\left( \frac{c_{+}}p\right) ^2}-2%
\sqrt{1+\left( \frac cp\right) ^2}\right]
\end{equation}
and for $p^2>>c_{\mp }^2,c^2$, we obtain 
\[
\frac V{2\pi ^2}\frac 12\sum_{l=0}^\infty \int_0^\infty dpp^3\left[ 1+\frac 1%
2\left( \frac{c_{-}}p\right) ^2-\frac 18\left( \frac{c_{-}}p\right) ^4+1+%
\frac 12\left( \frac{c_{+}}p\right) ^2-\frac 18\left( \frac{c_{+}}p\right)
^4\right. 
\]
\begin{equation}
\left. -2-\left( \frac cp\right) ^2-\frac 14\left( \frac cp\right) ^4\right]
=-\frac V{2\pi ^2}\frac{c^4}8\int_0^\infty \frac{dp}p.
\end{equation}
We will use a cut-off $\Lambda $ to keep under control the $UV$ divergence%
\footnote{%
It is known that at one-loop level Gravity is renormalizable only in flat
space. In a dimensional regularization scheme its contribution to the action
is, on shell, proportional to the Euler character of the manifold that is
nonzero for the Schwarzschild instanton. Although in our approach we are
working with sections of the original manifold to deal with these
divergences one must introduce a regulator that indeed appears in the
contribution of energy density.} 
\begin{equation}
\int_0^\infty \frac{dp}p\sim \int_0^{\frac \Lambda c}\frac{dx}x\sim \ln
\left( \frac \Lambda c\right) .
\end{equation}
Thus $\Delta E\left( m\right) $ for high momenta becomes 
\begin{equation}
\Delta E\left( m\right) \sim -\frac V{2\pi ^2}\frac{c^4}{16}\ln \left( \frac{%
\Lambda ^2}{c^2}\right) =-\frac V{2\pi ^2}\left( \frac{3m}{r_0^3}\right) ^2%
\frac 1{16}\ln \left( \frac{r_0^3\Lambda ^2}{3m}\right) .  \label{d10}
\end{equation}
At this point we can compute the total energy, namely the classical
contribution plus the quantum correction up to second order. Recalling the
definition of asymptotic energy for an asymptotically flat background, such
as the Schwarzschild one gets, 
\begin{equation}
M-\frac V{2\pi ^2}\left( \frac{3m}{r_0^3}\right) ^2\frac 1{16}\ln \left( 
\frac{r_0^3\Lambda ^2}{3m}\right) =M-\frac V{2\pi ^2}\left( \frac{3MG}{r_0^3}%
\right) ^2\frac 1{16}\ln \left( \frac{r_0^3\Lambda ^2}{3MG}\right)
\end{equation}
One can observe that 
\begin{equation}
\Delta E\left( m\right) \rightarrow \infty \text{ when }m\rightarrow 0\text{%
, for }r_0=2m=2GM
\end{equation}
and 
\begin{equation}
\Delta E\left( m\right) \rightarrow 0\text{ when }m\rightarrow 0\text{, for }%
r_0\neq 2m=2GM.
\end{equation}
{\bf Remark } We would like to explain the reasons that support the results
of formula $\left( \ref{d10}\right) $. In that formula we introduced a
particular value of the radius, which behaves as a regulator with respect to
the horizon approach of the potential. The meaning of this particular value
is related to the necessity of explaining the dynamical origin of black hole
entropy by the entanglement entropy mechanism and by the so-called ``{\it %
brick wall model}'' \cite{t Hooft,FroNov}. Indeed, the same mechanism is
present when one has to regularize entropy by imposing a kind of cut-off,
which in coordinate space means $r_0>2m$. In fact, $r_0$ can be seen as $%
2m+h $, where $h$ assumes the same meaning of Ref.\cite{t Hooft}. However,
to explicitly relate this quantity we have to compare the Bekenstein-Hawking
entropy with the result deriving from the evaluation of the free energy for
gravitons, in this case. The only difference from the original calculation
is the spin contribution not present for scalar fields.

\section{Summary and Conclusions}

\label{p5}

We started from the problem of defining quantum corrections (semiclassical)
to a gravitational energy. By means of a variational approach with Gaussian
wave functionals an attempt to calculate such a correction was made. Despite
the constraint equations, this calculation is based on an extension of the
subtraction procedure involving volume terms in the semiclassical regime.
Excitations coming from boundary terms have been neglected to avoid the
unphysical situation of having contributions deriving from infinity. In this
context the extended subtraction procedure corresponds to the difference
between zero point energies calculated in an asymptotically flat curved
background referring to a flat background. This procedure eliminates the UV
divergence of the free gravitons leaving the contribution of the curved
background related to an {\it imposed by hand} UV cut-off. A strong analogy
with the Casimir vacuum energy calculation is revealed, opening the
possibility of understanding several configurations and their relationship
with the vacuum stability. Indeed, this apparatus can be applied also to the
Schwarzschild-deSitter background which asymptotically approaches the
deSitter space and so on. Although this evaluation has to be completed with
a careful study of the spin-two operator, we can conclude that the
variational approach for the computation of quantum corrections
(semiclassical) to a classical energy can be thought of as a tool for
testing zero point energy (Casimir-like energy) in a complicated theory such
as Einstein gravity

\section{Acknowledgments}

I wish to thank G. Esposito, V.Frolov, D.L. Rapoport for our helpful
discussions. I also thank S. Liberati and B. Jensen for their suggestions on
how to justify the horizon approach.

\appendix

\section{Conventions}

\label{p6}

\begin{enumerate}
\item  Riemann tensor, Ricci tensor and the Scalar Curvature in $3D$%
\[
R_{ijm}^{l}=\Gamma _{mi,j}^{l}-\Gamma _{ji,m}^{l}+\Gamma _{ja}^{l}\Gamma
_{mi}^{a}-\Gamma _{ma}^{l}\Gamma _{ji}^{a}\text{ \ Riemann tensor}
\]
Because of the vanishing of the Weyl tensor in $3D$, that is $C_{ijm}^{l}=0$%
, the Riemann tensor is completely determined by the Ricci tensor 
\[
R_{lijm}^{{}}=g_{lj}R_{im}^{{}}-g_{lm}R_{ij}^{{}}-g_{ij}R_{lm}^{{}}+g_{im}R_{lj}^{{}}
\]
\[
R_{im}=R_{ilm}^{l}\text{ \ Ricci tensor}
\]
\[
R=g_{{}}^{lj}R_{lj}^{{}}\text{ \ Scalar curvature}
\]
\end{enumerate}

\section{The Kinetic Term}

\label{p7}

The Schr\"{o}dinger picture representation of the kinetic term is 
\begin{equation}
G_{ijkl}\pi ^{ij}\pi ^{kl}=G_{ijkl}\left( -\frac{\delta ^2}{\delta
h_{ij}\left( x\right) \delta h_{kl}\left( x\right) }\right) .
\end{equation}
We have to apply this quantity to the gaussian wave functional $\left| \Psi
\right\rangle $. This means that 
\[
\pi ^{ij}\left( x\right) \pi ^{kl}\left( x\right) \left| \Psi \right\rangle
=-\frac{\delta ^2\Psi \left[ h\right] }{\delta h_{ij}\left( x\right) \delta
h_{kl}\left( x\right) } 
\]
\[
=\frac 12K^{-1\left( kl\right) \left( ij\right) }\left( x,x\right) \left( 
\sqrt{g\left( x\right) }\right) ^2\Psi \left[ h\right] 
\]
\[
-\frac 14\int d^3y^{\prime }d^3y^{\prime \prime }\left( \sqrt{g\left(
x\right) }\right) ^2\sqrt{g\left( y^{\prime }\right) }\sqrt{g\left(
y^{\prime \prime }\right) }K^{-1\left( kl\right) \left( k^{\prime }l^{\prime
}\right) }\left( x,y^{\prime }\right) h_{k^{\prime }l^{\prime }}\left(
y^{\prime }\right) 
\]
\begin{equation}
\cdot K^{-1\left( ij\right) \left( k^{\prime \prime }l^{\prime \prime
}\right) }\left( x,y^{\prime \prime }\right) h_{k^{\prime \prime }l^{\prime
\prime }}\left( y^{\prime \prime }\right) \Psi \left[ h\right] .  \label{ab1}
\end{equation}
By functional integrating 
\begin{equation}
\left\langle \Psi \left| h_{k^{\prime }l^{\prime }}\left( y^{\prime }\right)
h_{k^{\prime \prime }l^{\prime \prime }}\left( y^{\prime \prime }\right)
\right| \Psi \right\rangle =K_{\left( k^{\prime }l^{\prime }\right) \left(
k^{\prime \prime }l^{\prime \prime }\right) }\left( y^{\prime },y^{\prime
\prime }\right) \left\langle \Psi |\Psi \right\rangle .
\end{equation}
Thus 
\[
\left\langle \Psi \left| \pi ^{ij}\left( x\right) \pi ^{kl}\left( x\right)
\right| \Psi \right\rangle 
\]
becomes 
\[
\frac 12K^{-1\left( kl\right) \left( ij\right) }\left( x,x\right) \left( 
\sqrt{g\left( x\right) }\right) ^2 
\]
\[
-\frac 14\int d^3y^{\prime }d^3y^{\prime \prime }\left( \sqrt{g\left(
x\right) }\right) ^2\sqrt{g\left( y^{\prime }\right) }\sqrt{g\left(
y^{\prime \prime }\right) }K^{-1\left( kl\right) \left( k^{\prime }l^{\prime
}\right) }\left( x,y^{\prime }\right) K^{-1\left( ij\right) \left( k^{\prime
\prime }l^{\prime \prime }\right) }\left( x,y^{\prime \prime }\right) 
\]
\[
K_{\left( k^{\prime }l^{\prime }\right) \left( k^{\prime \prime }l^{\prime
\prime }\right) }\left( y^{\prime },y^{\prime \prime }\right) \left\langle
\Psi |\Psi \right\rangle 
\]
\begin{equation}
=\frac 14K^{-1\left( kl\right) \left( ij\right) }\left( x,x\right) \left( 
\sqrt{g\left( x\right) }\right) ^2\left\langle \Psi |\Psi \right\rangle .
\end{equation}
Then the expectation value of the kinetic term, with the Planck length
reinserted, is 
\begin{equation}
\left\langle T\right\rangle =\frac 1{4l_p^2}\int d^3x\sqrt{g}\left(
G_{ijkl}K^{-1\left( kl\right) \left( ij\right) }\left( x,x\right) \right) ,
\label{ab2}
\end{equation}

\end{document}